\documentclass[9pt,twocolumn,twoside]{opticajnl}

\journal{opticajournal}
\setboolean{shortarticle}{true}
\setboolean{displaycopyright}{false}

\usepackage{graphicx}
\usepackage{amsmath}
\usepackage{lineno}
\linenumbers
\nolinenumbers
\title{Propagation-tunable fractional Stokes skyrmions in conical refraction}

\author[1,*]{Nawal Mohamed}
\author[1]{Fatima Khanom}
\author[1]{Diana Galiakhmetova}
\author[1]{Shakti Singh}
\author[2]{Valentin Mylnikov}
\author[2]{Grigorii Sokolovskii}
\author[1]{Edik Rafailov}
\author[1]{Igor Meglinski}

\affil[1]{Aston Institute of `Photonic Technologies,
Aston University, Birmingham, United Kingdom}
\affil[2]{A.F. Ioffe Physico-Technical Institute, Russian Academy of Sciences, Saint Petersburg, Russia}
\affil[*]{Correspondence: mohamen7@aston.ac.uk~~~~and~/~or~~~~i.meglinski@aston.ac.uk}

\begin{abstract}
Conical refraction in a biaxial crystal generates Stokes textures whose finite-domain skyrmion number takes fractional values tunable through free-space propagation, while the underlying azimuthal winding remains unchanged. For a circularly polarised Gaussian input, the skyrmion number is governed by the boundary Stokes parameter, $N_\mathrm{sk}=[1-s_3(R_\mathrm{cut},Z)]/2$. Six-state polarimetry reveals fractional values of 0.523, 0.565 and 0.266 in the Lloyd, Poggendorff and Raman regions, respectively, in close agreement with Belsky--Khapalyuk--Berry predictions of 0.530, 0.550 and 0.250, and reproduces the predicted longitudinal evolution. Conical refraction thus provides a passive route to propagation-tunable fractional Stokes skyrmions, in which continuously evolving polarisation geometry coexists with a fixed topological winding.
\end{abstract}

\begin{document}
\maketitle

Optical skyrmions have emerged as a class of topologically structured electromagnetic fields in which spatially varying polarisation states form nontrivial mappings onto the Poincar\'e sphere~\cite{Shen2024Skyrmions,GutierrezCuevas2021Optical}. Their complex vectorial textures and topological properties have stimulated interest in optical information encoding, spin--orbit photonics, imaging and precision metrology. The skyrmion number, defined through the spatial distribution of the normalised Stokes vector, provides a quantitative description of these mappings~\cite{Cisowski2023Rational,McWilliam2023Topological}.

Considerable progress has been made in generating and manipulating optical skyrmions through engineered superpositions of orthogonally polarised spatial modes, interfering optical fields and spin--orbit-coupled photonic structures. Marco \emph{et al.} demonstrated propagation-invariant optical meron lattices formed through plane-wave interference~\cite{Marco2024Meron}, while recent investigations have extended the accessible topological configurations to non-integer skyrmion numbers. In particular, Cao \emph{et al.} experimentally generated fractional skyrmions using vectorial superpositions of fractional orbital angular momentum (OAM) modes~\cite{Cao2026Fractional}. Furthermore, propagation-dependent topological transitions have been demonstrated in engineered vector beams~\cite{Fan2025Propagation,Zhen2026Reconfiguring}, establishing longitudinal evolution as an additional degree of freedom for controlling spatial polarisation topology.

Despite these advances, the evolution of finite-domain Stokes topology arising from the intrinsic spin--orbit transformation of conical refraction (CR) remains comparatively underexplored experimentally. In biaxial crystals, CR generates coupled polarisation components whose spatial amplitude and phase evolve through the characteristic Lloyd, Poggendorff and Raman configurations~\cite{Berry2006Conical,Turpin2016Conical,Sokolovskii2013}. Previous studies have investigated their phase and OAM properties~\cite{Galiakhmetova2025Topological}; however, the relationship between propagation-dependent Stokes textures, non-integer skyrmion numbers and finite-domain boundary polarisation has not been systematically established for these configurations. This distinction is essential because fractional values measured over a finite transverse domain do not necessarily represent globally quantised topological charges.

Here, we experimentally investigate the propagation-dependent evolution of finite-domain Stokes skyrmion numbers generated by CR. Six-state polarimetry is employed to reconstruct the polarisation textures in the Lloyd, Poggendorff and Raman regions, with the resulting skyrmion numbers compared against Belsky--Khapalyuk--Berry (BKB) predictions. Their longitudinal evolution is further interpreted through the boundary Stokes component and modelled electromagnetic energy-flow trajectories. The results establish CR as a passive platform for accessing distinct finite-domain polarisation configurations through propagation alone.


A circularly polarised Gaussian beam incident along an optic axis of a biaxial crystal undergoes a spin-orbit transformation. The paraxial Belsky--Khapalyuk--Berry (BKB) description contains two coupled radial amplitudes, $B_0$ and $B_1$, corresponding to azimuthally independent and vortex-bearing field components~\cite{Berry2006Conical,Turpin2016Conical,Mylnikov2020}. In the circular-basis convention implemented in our numerical model,
\begin{equation}
\mathbf{E}(\rho,\varphi,Z)=
B_0(\rho,Z)\mathbf{e}_{+}
+B_1(\rho,Z)e^{i\varphi}\mathbf{e}_{-},
\label{eq:BKB}
\end{equation}
where $\rho=r/w_0$, $\varphi$ is the transverse azimuth, $Z=z/z_R$ is the normalised propagation distance, and $\mathbf e_{\pm}$ are circular-basis vectors. The amplitudes follow BKB diffraction integrals involving Bessel functions of orders zero and one, consistent with earlier experimental observations of spin-orbit coupling and vortex generation in CR~\cite{Peet2014}. Their relative phases and magnitudes change during propagation, producing the Lloyd, Poggendorff and Raman field configurations.

The spatial polarisation structure emerges from the coherent superposition of the two circularly polarised field components. The azimuthally independent contribution $B_0$ and vortex-bearing contribution $B_1e^{i\varphi}$ possess different radial amplitude and phase distributions, which  evolve during propagation. Their relative amplitudes determine the local circular polarisation content, whereas their relative phase governs the orientation of the transverse polarisation structure. Consequently, the normalised Stokes vector develops a spatially varying distribution across the beam.

The normalised Stokes field is $\mathbf{s}=(s_1,s_2,s_3)$, with $s_i=S_i/S_0$. Its finite-domain skyrmion number is
\begin{equation}
N_{\mathrm{sk}}(Z)=\frac{1}{4\pi} \iint_{\Omega}\mathbf{s}\cdot  \left(\frac{\partial\mathbf{s}}{\partial x} \times\frac{\partial\mathbf{s}}{\partial y}\right)\,dx\,dy,
\label{eq:Nsk}
\end{equation}
where $\Omega$ is a fixed circular disk of radius $\rho=R_{\mathrm{cut}}=11$ in normalised coordinates, used for the theoretical and experimental comparisons. For the ideal axisymmetric model and its adopted winding and sign convention, the integral reduces to
\begin{equation}
N_{\mathrm{sk}}(Z)=\frac12
\left[s_3(0,Z)-s_3(R_{\mathrm{cut}},Z)\right].
\label{eq:boundary}
\end{equation}

For the adopted circular-basis convention, the Stokes parameter $s_3$ describes the relative intensities of the two circularly polarised components. In the ideal axisymmetric BKB field, the vortex-bearing contribution vanishes at the beam centre, giving $s_3(0,Z)=1$ wherever the central intensity is nonzero. The finite-domain skyrmion number is therefore governed by the polarisation state at the fixed outer boundary.

Since this boundary state is not generally constrained to a single polarisation pole, the integrated Stokes mapping need not correspond to an integer-valued global topological invariant. The resulting non-integer values reflect the finite observation domain and its evolving boundary polarisation, rather than requiring fractional azimuthal winding.

\begin{figure}[!h]
\centering
\includegraphics[width=\columnwidth]{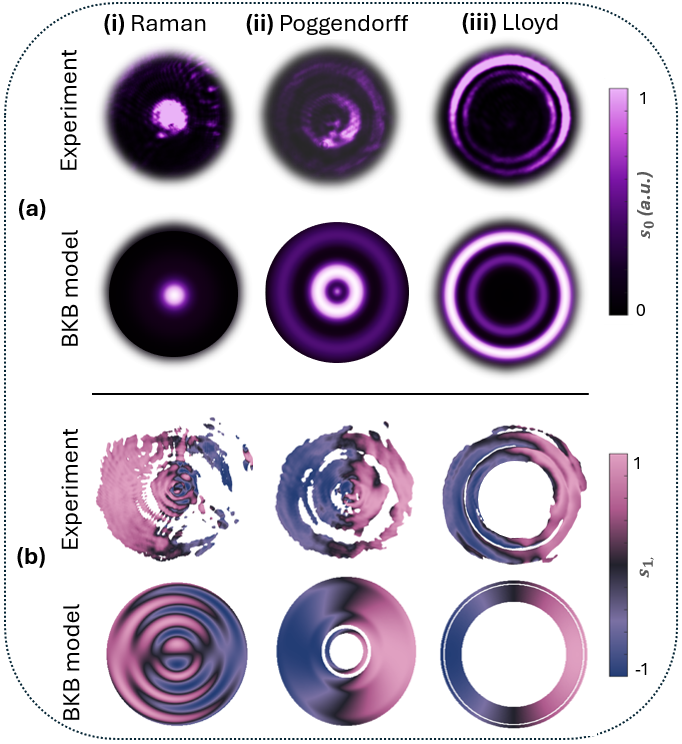}
\caption{Experimental and theoretical CR fields in the Raman, Poggendorff and Lloyd regions (left to right). (a) Experimental and BKB intensity distributions ($S_0$). (b) Corresponding normalised Stokes-component distributions ($s_1=S_1/S_0$).}
\label{fig:stokes_comparison}
\end{figure}

CR was generated in a 3~at\% $Nd:KGd(WO_4)_2$ biaxial crystal illuminated along an optic axis by a circularly polarised $633~nm$ $He-Ne$ Gaussian beam (linear polariser and quarter-wave plate) focused by a $25~mm$ lens, as described previously~\cite{Galiakhmetova2025Topological,Sokolovskii2013}. A translatable $150~mm$ lens selected the observation plane, and the field was imaged onto a CMOS camera (Thorlabs DCC3240M, $1280\times1024$ pixels). Spatially resolved Stokes parameters were obtained by six-state polarimetry~\cite{Schaefer2007}, from intensity images recorded behind a quarter-wave plate and analyser for horizontal, vertical, diagonal, antidiagonal, right- and left-circular projections, with $S_0=I_H+I_V$,
$S_1=I_H-I_V$, $S_2=I_D-I_A$ and $S_3=I_R-I_L$. The images were registered on a common grid, smoothed and restricted to an intensity-based region of interest, and the derivatives in (\ref{eq:Nsk}) were evaluated numerically on the camera grid. Six-state polarimetry gave more reproducible Stokes maps than interferometric reconstruction of the orthogonal field components, and complements earlier use of CR for polarisation metrology~\cite{Peinado2013Conical}. Figure~\ref{fig:stokes_comparison} compares the measured intensity and $s_1$ distributions with the BKB model.

\begin{figure}[!t]
\centering
\includegraphics[width=\columnwidth]{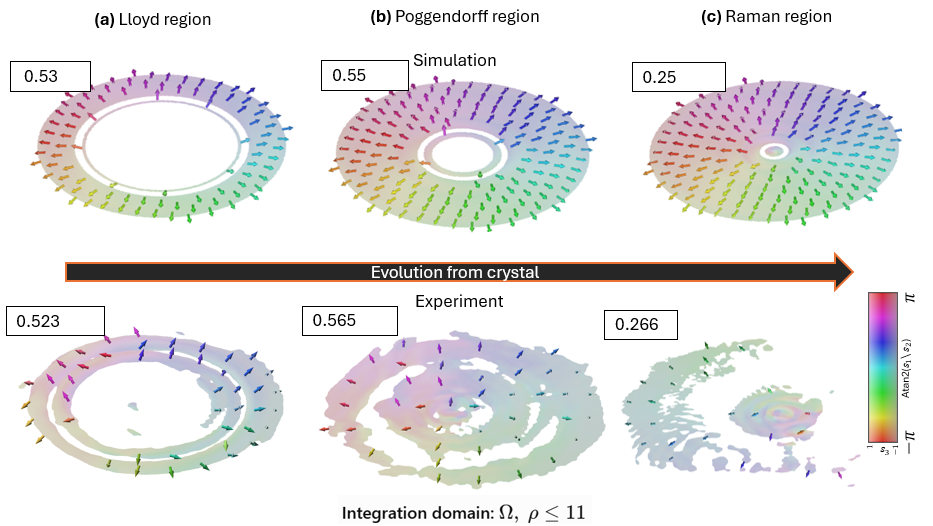}
\caption{Fractional Stokes textures in the Lloyd, Poggendorff and Raman regions. BKB simulations (upper row) are compared with experimental polarisation reconstructions (lower row). Colour encodes the polarisation-ellipse orientation through $\mathrm{atan2}(s_2,s_1)=2\psi$, as in Ref.~\cite{Cao2026Fractional}; arrows show the full Stokes vector, with their out-of-plane component giving $s_3$. The finite-domain skyrmion numbers are indicated for each configuration. The integration domain $\Omega$ is bounded by $\rho=R_{\mathrm{cut}}=11$ in normalised transverse coordinates.
} \label{fig:textures} 
\end{figure}

\begin{figure*}[!h]
\centering
\includegraphics[width=\textwidth]{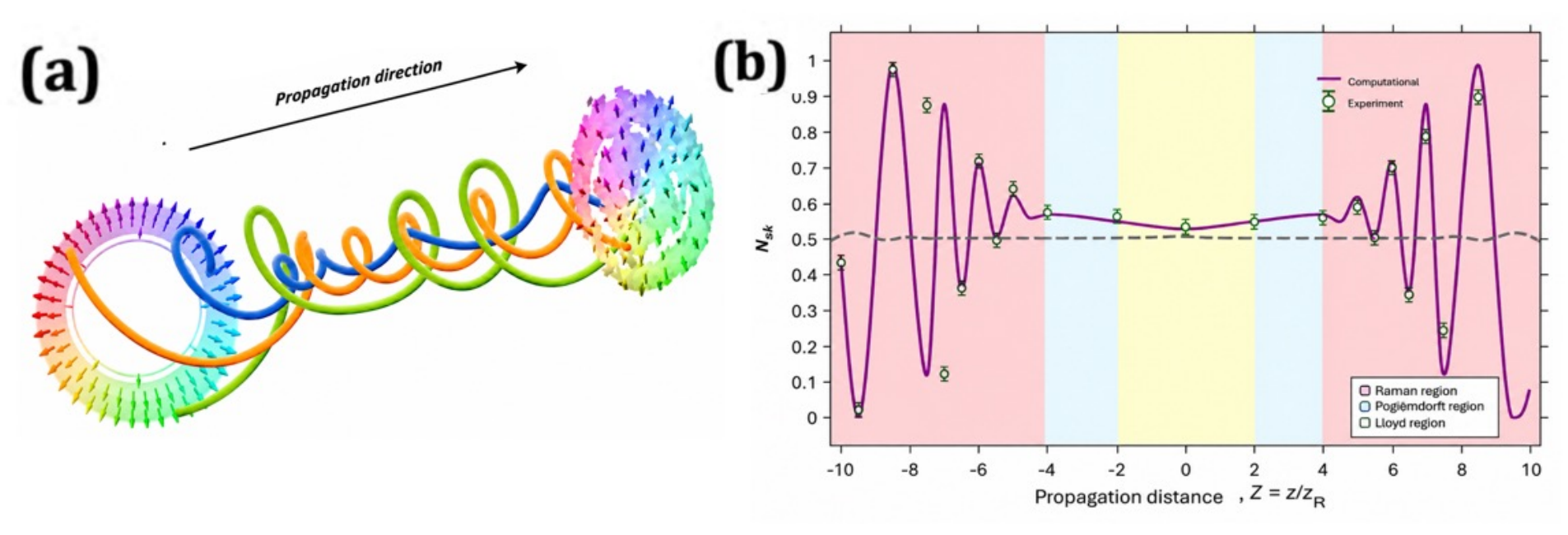}
\caption{Propagation-dependent evolution of finite-domain Stokes skyrmion numbers in CR. (a) Schematic illustration of energy-flow trajectories (integral curves of the time-averaged Poynting vector, Eq.(\ref{eq:poynting})) from the Lloyd plane ($Z=0$) to the Raman region ($Z=15$), shown with the transverse Stokes-vector textures at the initial and final planes. The trajectories wind in three dimensions while the transverse texture remains organised about the axis; its azimuthal winding is fixed by the field symmetry. (b) Corresponding evolution of the Stokes skyrmion number, comparing experimental measurements with the continuous theoretical model across the propagation range. The dashed line marks $N_\mathrm{sk}=1/2$, the value obtained when the domain boundary is placed at the illuminated periphery of the beam. The shaded regions indicate the Lloyd, Poggendorff, and Raman regimes.}
\label{fig:propagation} 
\end{figure*}

Figure~\ref{fig:textures} compares BKB (upper row) and experimental (lower row) Stokes-vector textures across the three CR regions. Their distinct spatial organisations reflect the coupled polarisation components of the conically refracted field~\cite{Turpin2016Conical,Galiakhmetova2025Topological}.

The Lloyd region exhibits an approximately annular distribution with systematic azimuthal variation of the Stokes vectors. The colour encodes the Stokes longitude $\Phi=\mathrm{atan2}(s_2,s_1)=2\psi$, i.e.\ the azimuthal coordinate of $\mathbf{s}$ on the Poincar\'e sphere, while $s_3$ sets its latitude. For the BKB field $\Phi=\arg(E_+E_-^{*})=-\varphi+\arg(B_0B_1^{*})$, so on any concentric circle the colour completes exactly one cycle and the ellipse orientation $\psi$ rotates by $\pi$: the beam axis, where $s_3=1$, is a polarisation C-point of index $1/2$~\cite{BerryDennis2001}. The magnitude of this azimuthal winding is unity, with its sign determined by the adopted Stokes and circular-basis conventions; the winding remains fixed during propagation. The non-integer skyrmion number has a geometric origin: for the ideal axisymmetric field, each concentric circle maps onto a circle of latitude on the Poincar\'e sphere, and $N_\mathrm{sk}$ is determined by the enclosed solid angle divided by $4\pi$, as expressed by (\ref{eq:boundary}). On the Lloyd ring the polarisation is close to linear ($s_3\approx0$), so the cap is approximately one hemisphere and $N_\mathrm{sk}\approx1/2$; propagation leaves the colour winding unchanged and alters only this boundary latitude.
Its experimental skyrmion number, $N_{\mathrm{sk}}=0.523$, differs from the BKB value of $0.530$ by $0.007$. In the Poggendorff region, concentric intensity structure is accompanied by more intricate radial polarisation organisation; the experimental and BKB values are $0.565$ and $0.550$, respectively. The Raman configuration displays distinct central and peripheral features and a less continuous experimental reconstruction, giving $0.266$ compared with the predicted $0.250$.

The absolute differences of $0.007$, $0.015$ and $0.016$ establish close agreement for these three integrated values, although local map discrepancies remain. In the Raman region, low-intensity pixels and discontinuous valid-data areas may particularly affect the stability of the experimentally evaluated integral.
Polarisation winding and boundary conditions are essential to the interpretation of optical skyrmion numbers~\cite{McWilliam2023Topological}; agreement of integrated values alone does not demonstrate identical pointwise Stokes fields. Since the circular integration radius is fixed at $R_{\mathrm{cut}}=11$, the observed differences describe changes in the polarisation mapping within a common domain rather than changes in aperture size.

The longitudinal evolution of the finite-domain Stokes skyrmion number is presented in Fig.~\ref{fig:propagation}(b),  where experimental measurements are compared with the continuous BKB prediction over the normalised propagation interval $-10 \leq Z \leq 10$. The theoretical curve was obtained by computing $B_0$ and $B_1$ from Eq.~(\ref{eq:BKB}) at each propagation plane and evaluating Eq.~(4) from $s_3(R_\mathrm{cut},Z)$, whereas the experimental points were obtained independently by evaluating Eq.~(\ref{eq:Nsk}) directly on the measured Stokes maps. Error bars represent one standard deviation of repeated measurements. The model reveals distinct propagation-dependent behaviour across the characteristic CR regions. In the central Lloyd region and neighbouring Poggendorff region, the calculated skyrmion number exhibits comparatively gradual variations within a restricted positive range. In contrast, the outer Raman regions display more pronounced oscillations, reflecting changes in the spatial polarisation structure during propagation. The experimental measurements reproduce the principal longitudinal behaviour, although deviations from the ideal prediction are observed at individual positions. The representative configurations in Fig.~\ref{fig:textures} therefore correspond to selected propagation planes within this continuous evolution.

The physical origin of the calculated finite-domain evolution can be examined through the boundary-polarisation contribution. For the ideal axisymmetric BKB field, the vortex-bearing component vanishes at the beam centre, giving $s_3(0,Z)=1$ wherever the central intensity is nonzero. Consequently, Eq.~(\ref{eq:boundary}) becomes
\begin{equation}
N_{\mathrm{sk}}(Z)=
\frac{1-s_3(R_{\mathrm{cut}},Z)}{2}.
\label{eq:boundary_evolution}
\end{equation}
The corresponding boundary contribution, $Q_\mathrm{boundary}(Z)=-s_3(R_\mathrm{cut},Z)/2$, follows directly from the BKB curve in Fig.~\ref{fig:propagation}(b), since $N_\mathrm{sk}(Z)=1/2+Q_\mathrm{boundary}(Z)$. Its variation demonstrates how
the polarisation state at the fixed transverse boundary influences the modelled skyrmion number. This contribution represents a decomposition of the same BKB prediction rather than an independent experimental measurement.

A complementary description of the evolving CR field is provided by electromagnetic energy-flow trajectories, illustrated schematically in Fig.~\ref{fig:propagation}(a). Energy-flow trajectories are defined as integral curves of the time-averaged Poynting vector, $dx/dz=P_x/P_z$, $dy/dz=P_y/P_z$~\cite{BerryMcDonald2008,Berry2009},
\begin{equation}
\mathbf{P}=\frac{1}{2}
\operatorname{Re}
\left[\mathbf{E}\times\mathbf{H}^{*}\right],
\label{eq:poynting}
\end{equation}
where $\mathbf{E}$ and $\mathbf{H}$ denote the complex electric and magnetic field amplitudes, with $\mathbf{H}=(i\omega\mu_0)^{-1}\nabla\times\mathbf{E}$ evaluated from the BKB field, including its first-order longitudinal component. The trajectories in Fig.~\ref{fig:propagation}(a) show that the conically refracted field, which carries OAM~\cite{Berry2005CRAM}, can follow complex three-dimensional paths while retaining an organised transverse Stokes texture at each observation plane. This behaviour parallels the OAM phase-memory mechanism, in which the azimuthal phase is fixed by the rotational symmetry of the source field rather than by individual trajectories~\cite{Meglinski2024LSA,Khanom2024,Eshaghi2022}, and is sampled along spiralling energy-flow paths~\cite{BerryMcDonald2008}. In the axisymmetric BKB field, the corresponding symmetry-fixed quantities are the unit azimuthal winding of the vortex-bearing component, $B_1 e^{i\varphi}$, and the on-axis state $s_3(0,Z)=1$. Propagation reshapes the radial amplitudes and relative phase of $B_0$ and $B_1$, and therefore the boundary polarisation that sets $N_\mathrm{sk}$ through Eq.~(\ref{eq:boundary_evolution}), but does not alter this winding. The trajectories thus illustrate how energy is redistributed around a symmetry-preserved vector structure; they are not a direct measure of skyrmion charge or evidence of topological protection of $N_\mathrm{sk}$. The spiralling energy-flow trajectories reflect the azimuthal circulation associated with the OAM-bearing component, whereas the fractional skyrmion number arises from the propagation-dependent redistribution of the coupled polarisation components and the resulting boundary Stokes state.


The experimental observations demonstrate that CR provides a mechanism for accessing distinct finite-domain Stokes skyrmionic textures through longitudinal propagation. The agreement between the experimental skyrmion numbers and BKB predictions across the Lloyd, Poggendorff and Raman regions indicates that the model captures the principal integrated polarisation characteristics, despite local differences in the reconstructed Stokes-vector distributions.

The fractional values observed here are distinct from those generated through engineered fractional OAM superpositions~\cite{Cao2026Fractional}. There, the non-integer value originates from a branch cut in the azimuthal phase, which produces an abrupt azimuthal jump of the polarisation, incomplete azimuthal coverage of the Poincar\'e sphere and abrupt transitions toward integer values on propagation. In CR the Stokes longitude winds continuously, with no branch cut, and the coverage is limited instead in latitude by the boundary state $s_3(R_\mathrm{cut},Z)$, which evolves smoothly with propagation. In the present axisymmetric BKB description, the azimuthal winding remains fixed, while the finite-domain skyrmion number varies with the polarisation state at the integration boundary. This distinction is consistent with the importance of boundary conditions in determining optical skyrmion topology~\cite{Shen2024Skyrmions,McWilliam2023Topological}. Consequently, the observed variations should be interpreted as changes in the finite-domain Stokes mapping, rather than necessarily as transitions between globally quantised topological states. When the domain boundary is placed at the illuminated periphery of the beam, where the CR field is linearly polarised, $N_\mathrm{sk}\approx1/2$ at all $Z$, i.e. a meron-like texture; the propagation dependence in Fig.~\ref{fig:propagation}(b) therefore reflects the evolution of the field relative to a fixed observation domain, which is the relevant quantity for any measurement with finite aperture.

Propagation-dependent transformations have previously been demonstrated using engineered vector beams~\cite{Fan2025Propagation,Zhen2026Reconfiguring}. In contrast, the present approach exploits the intrinsic spin--orbit coupling of a biaxial crystal. The coupled BKB field components evolve naturally during propagation, producing different polarisation configurations without external spatial beam shaping.

The experimental interpretation remains subject to finite spatial resolution, reduced polarimetric reliability in low-intensity regions, and sensitivity of the skyrmion integral to spatial derivatives. Furthermore, the boundary-polarisation relation is derived for the ideal axisymmetric model and does not independently validate the experimental topology.


In summary, we demonstrate that conical refraction enables propagation-tunable fractional Stokes skyrmion numbers, while the underlying azimuthal winding remains fixed. Six-state polarimetry shows close agreement with Belsky--Khapalyuk--Berry predictions across the Lloyd, Poggendorff and Raman regions. The observed evolution arises from changes in the polarisation state at a fixed transverse boundary, rather than from transitions between quantised topological states. These findings establish conical refraction as a passive platform for controlling finite-domain polarisation geometry and illustrate how continuously evolving Stokes textures can coexist with a fixed topological winding.

\begin{backmatter}
\bmsection{Funding}
The study was supported by the European Union’s Horizon Europe Research and Innovation Programme, OPTIPATH project, Grant Agreement No. 101185769. Authors also acknowledge the support from COST Action CA23125 - The mETamaterial formalism approach to recognize cAncer (TETRA). 


\bmsection{Disclosures}
The authors declare no conflicts of interest.

\bmsection{Data availability}
Data underlying the results presented in this
paper are not publicly available at this time
but may be obtained from the authors upon
reasonable request.
\end{backmatter}

\bibliography{sample}
\end{document}